\documentclass[fleqn,usenatbib,usedcolumn]{mnras}

\usepackage{newtxtext,newtxmath}
\usepackage[T1]{fontenc}
\usepackage{ae,aecompl}
\usepackage{graphicx}
\usepackage{amsmath}
\usepackage{amssymb}
\usepackage{color}
\usepackage{times}
\usepackage{multirow}
\usepackage{booktabs}
\usepackage{adjustbox}
\usepackage{hyperref}

\definecolor{RoyalBlue}{rgb}{0.25,0.41,0.88}
\definecolor{Red}{rgb}{0.61,0.12, 0.14}

\volume{{\rm in press}}         % good for ArXiv

\title[Protolunar disc masses]{Upper limits on protolunar disc masses
  using ALMA observations of directly-imaged exoplanets }

\author[S. Perez et al.]{Sebasti\'an P\'erez,$^{1,2}$\thanks{Corresponding author 1: sebastian.astrophysics@gmail.com}
  Sebasti\'an Marino,$^3$\thanks{Corresponding author 2: sebastian.marino.estay@gmail.com}
  Simon Casassus,$^1$
  Cl\'ement Baruteau,$^{4}$
  \newauthor 
  Alice Zurlo,$^{5,6}$
  Christian Flores$^{7}$
  and Gael Chauvin$^{8,9}$
  \\
  $^{1}$Universidad de Santiago de Chile, Av. Ecuador 3659, Santiago\\
  $^{2}$Departamento de Astronom\'ia, Universidad de Chile, Casilla 36-D, Santiago\\
  $^{3}$Max Planck Institute for Astronomy, K\"onigstuhl 17, 69117 Heidelberg, Germany\\
  $^{4}$IRAP, Universit\'e de Toulouse, CNRS, UPS, Toulouse, France\\
  $^{5}$N\'ucleo de Astronom\'ia, Facultad de Ingenier\'ia y Ciencias, Universidad Diego Portales, Av. Ejercito 441, Santiago, Chile\\
  $^{6}$Escuela de Ingenier\'ia Industrial, Facultad de Ingenier\'ia y Ciencias, Universidad Diego Portales, Av. Ejercito 441, Santiago, Chile \\
  $^{7}$Institute for Astronomy, University of Hawaii at Manoa, 640 N. Aohoku Place, Hilo, HI 96720, USA\\
  $^{8}$Unidad Mixta Internacional Franco-Chilena de Astronom\'ia, CNRS/INSU UMI 3386 \\
  $^{9}$Univ. Grenoble Alpes, CNRS, IPAG, F-38000 Grenoble, France
}

\pubyear{2019}

\begin{document}

\label{firstpage}
\pagerange{\pageref{firstpage}--\pageref{lastpage}}
\maketitle

\begin{abstract}
  The Solar System gas giants are each surrounded by many moons, with
  at least 50 prograde satellites thought to have formed from
  circumplanetary material. Just like the Sun is not the only star
  surrounded by planets, extrasolar gas giants are likely surrounded
  by satellite systems. Here, we report on ALMA observations of four
  <40 Myr old stars with directly-imaged companions: PZ Tel, AB Pic,
  51 Eri, and $\kappa$~And. Continuum emission at 1.3 mm is undetected
  for any of the systems. Since these are directly-imaged companions,
  there is knowledge of their temperatures, masses and
  locations. These allow for upper limits on the amount of
  circumplanetary dust to be derived from detailed radiative transfer
  models. These protolunar disc models consider two disc sizes: 0.4
  and 0.04 times the exoplanet's Hill radius. The former is
  representative of hydrodynamic simulations of circumplanetary discs
  while the latter a case with significant radial drift of solids. The
  more compact case is also motivated by the semi-major axis of
  Callisto, enclosing Jupiter's Galilean satellites. All upper limits
  fall below the expected amount of dust required to explain regular
  satellite systems ($\sim$10$^{-4}$ times the mass of their central
  planet). Upper limits are compared with viscous evolution and debris
  disc models. Our analysis suggests that the non detections can be
  interpreted as evidence of dust growth beyond metre sizes to form
  \textit{moonetesimals} in timescales $\lesssim10$~Myr. This sample
  increases by 50\% the number of ALMA non-detections of young
  companions available in the literature.
\end{abstract}

  %% The Solar System gas giants are each surrounded by many moons, with at least 50 prograde satellites thought to have formed from circumplanetary material. Just like the Sun is not the only star surrounded by planets, extrasolar gas giants are likely surrounded by satellite systems. Here, we report on ALMA observations of four <40 Myr old stars with directly-imaged companions: PZ Tel, AB Pic, 51 Eri, and κ And. Continuum emission at 1.3 mm is undetected for any of the systems. Since these are directly-imaged companions, there is knowledge of their temperatures, masses and locations. These allow for upper limits on the amount of circumplanetary dust to be derived from detailed radiative transfer models. These protolunar disc models consider two disc sizes: 0.4 and 0.04 times the exoplanet's Hill radius. The former is representative of hydrodynamic simulations of circumplanetary discs while the latter a case with significant radial drift of solids. The more compact case is also motivated by the semi-major axis of Callisto, enclosing Jupiter's Galilean satellites. All upper limits fall below the expected amount of dust required to explain regular satellite systems (10<sup>-4</sup> times the mass of their central planet). Upper limits are compared with viscous evolution and debris disc models. Our analysis suggests that the non detections can be interpreted as evidence of dust growth beyond metre sizes to form moonetesimals in timescales <10 Myr. This sample increases by 50% the number of ALMA non-detections of young companions available in the literature.


\section{Introduction}

\begin{table*}
  \centering
  \caption{Millimeter and sub-millimeter continuum observations of
    protolunar discs. Note that the stellar age and some of the
    exoplanet parameters (e.g. mass) are uncertain, hence the values
    used here serve simply as order of magnitude estimates.}
  \label{tab:obs}
  \begin{adjustbox}{max width=1.0\textwidth}
  \begin{tabular}{lccccccccccccr} 
    \toprule
    Exoplanet    & Age   & D$\dagger$ & \multicolumn{3}{c}{Stellar parameters} & \multicolumn{4}{c}{Exoplanet or companion parameters} & 3$\sigma$ limit & \multicolumn{2}{c}{$M_{\rm max}/M_{\rm p}$} & Refs.\\
    system       & (Myr) & (pc) & $M_\star$($M_{\odot}$) & $R_\star$ ($R_\odot$) &  $T_\star$ (K) & $a$ (au) &  $M_{\rm p}$ ($M_{\rm J}$) & $R_{\rm p}$ ($R_\odot$) & $T_{\rm p}$ (K)$^{\dagger\dagger}$ &  ($\mu$Jy) & $r_{\rm c}=0.4R_{\rm Hill}$ & $r_{\rm c}=0.04R_{\rm Hill}$ &\\
    %% \midrule
    %% \multicolumn{14}{c}{This work}\\
    \midrule
    PZ~Tel~b	  & 21 & 47  & 1.0  & 1.32 & 5250  & 25   & 38  & 0.25 & 2700 & 84  & 2.5$\times$10$^{-7}$ & 2.0$\times$10$^{-7}$ & 1,2\\ %% Maire2016A%26A...587A..56M,Pecaut+2012
    51~Eri~b	  & 21 & 30  & 1.75 & 1.5  & 7250  & 13   & 2   & 0.08 & 750  & 81  & 2.2$\times$10$^{-6}$ & -      & 3 \\ %% Macintosh+2015
    AB~Pic~b	  & 30 & 50  & 0.8  & 1.0  & 5000  & 260  & 13  & 0.15 & 1600 & 78  & 5.0$\times$10$^{-6}$ & 2.4$\times$10$^{-6}$ & 4,5\\ %% Bonnefoy+2010,Chauvin+2005
    $\kappa$~And~b& 47 & 50  & 2.5  & 2.3  & 11500 & 53   & 22  & 0.25 & 2040 & 180 & 9.0$\times$10$^{-7}$ & 1.4$\times$10$^{-6}$ & 6,7,8\\ %% Bonnefoy+2014,Hinkley+2013,Jones+2016
    \midrule
    \multicolumn{14}{c}{Previous 1.3 mm observations from \citet{Booth2016}}\\
    \midrule    
    HR~8799~b & 42 & 41  & 1.5  & 1.5  & 7250  & 70 & 5.8 & 0.12 & 870  & 48  & 1.1$\times$10$^{-6}$ & 1.5$\times$10$^{-6}$ & 9, 10, 24\\ %% Marois+2008,Booth+2016,Bonnefoy+2016
    HR~8799\{c,d,e\} & 42 & 41  & 1.5  & 1.5  & 7250  & \{43, 26, 15\} & 7.2 & 0.1 & 1100  & 48  & \{0.9,1.0,1.0\}$\times$10$^{-6}$ & \{1.2,1.4,1.6\}$\times$10$^{-6}$ & 9, 10, 24\\ %% Marois+2008,Booth+2016,Bonnefoy+2016
    %% HR~8799~b	 & 42 &	41  & 1.5  & 1.5  & 7250  & 69.5 & 5.8 & 0.12 &	870  & 48  & 1.1e-6 & 1.5e-6 & 9, 10, 24\\ %% Marois+2008,Booth+2016,Bonnefoy+2016
    %% HR~8799~c	 & 42 &	41  & 1.5  & 1.5  & 7250  & 43.3 & 7.2 & 0.10 &	1100 & 48  & 9.2e-7 & 1.2e-6 & 9, 10, 24\\ 
    %% HR~8799~d	 & 42 &	41  & 1.5  & 1.5  & 7250  & 25.6 & 7.2 & 0.10 &	1100 & 48  & 1.0e-6 & 1.4e-6 & 9, 10, 24\\ 
    %% HR~8799~e	 & 42 &	41  & 1.5  & 1.5  & 7250  & 15.4 & 7.2 & 0.10 &	1100 & 48  & 1.0e-6 & 1.6e-6 & 9, 10, 24\\ 
    \midrule
    \multicolumn{14}{c}{Previous 1.3 mm observations from \citet{Wu2017}}\\
    \midrule
    DH~Tau~b	 & 2  &	135 & 0.5  & 0.5  & 3600  & 317  & 15  & 0.25 &	2400 & 129 & 5.0$\times$10$^{-5}$ & 1.8$\times$10$^{-6}$ & 11,12,13\\ %% Wu+2017,Wolff+2017,Zhou+2014
    CT~Cha~b	 & 2  &	192 & 0.6  & 0.65 & 4000  & 512  & 20  & 0.22 &	2600 & 150 & 9.0$\times$10$^{-6}$ & 4.3$\times$10$^{-5}$ & 11,14\\ %% Wu+2017,+2015
    GSC6214-210~b& 10 &	109 & 0.9  & 1.0  & 4400  & 236  & 15  & 0.19 &	2200 & 90  & 6.0$\times$10$^{-6}$ & 2.5$\times$10$^{-6}$ & 11,15\\ %% Wu+2017,Bowler+2015
    1RXS1609~b	 & 10 &	140 & 0.85 & 1.3  & 4060  & 308  & 13  & 0.17 &	2000 & 90  & 2.0$\times$10$^{-5}$ & 7.0$\times$10$^{-6}$ & 11,16\\ %% Wu+2017,+2015ApJ...807L..13W
    GQ~Lup~b	 & 3  &	152 & 1.0  & 1.7  & 4300  & 109  & 25  & 0.33 &	2400 & 120 & 4.5$\times$10$^{-6}$ & 3.7$\times$10$^{-6}$ & 17 \\ %% Wu+2017ApJ...836..223W
    \midrule
    \multicolumn{14}{c}{Previous 0.89 mm observations from \citet{Ricci2017b}} \\
    \midrule
    2M1207~b	 & 8  &	52  & 0.023 & 0.36 & 2500 & 41   & 4   & 0.12 &	1100 & 78  & 1.5$\times$10$^{-6}$ & 1.8$\times$10$^{-6}$ & 18,19,20,21\\ %% Chauvin+04/05,Song06,Bowler2016,Ricci+17
    \midrule
    \multicolumn{14}{c}{Previous 1.3 mm observations from \citet{Matra2019}} \\
    \midrule
    $\beta$~Pic~b & 21 & 20  & 1.75  & 1.5  & 8250 & 9    & 11  & 0.15 & 1600 & 36  & 4.0$\times$10$^{-8}$ & 4.0$\times$10$^{-7}$ & 22,23 \\ %% lagrange+09/18,Bonnefoy2014,Matra+2018
    \midrule
    \multicolumn{14}{c}{Previous 1.3 mm observations from \citet{Su2017}} \\
    \midrule
    HD~95086~b	& 17 & 84   & 1.6   & 1.6  & 7550 & 62   & 4.4 & 0.12 & 1050 & 30  & 3.5$\times$10$^{-6}$ & 8.0$\times$10$^{-6}$ & 25,26\\ %% Su+2017,Zapata+2018,Rameau+2013
    \bottomrule
  \end{tabular}
  \end{adjustbox}
  \begin{flushleft}
    1. \citet{Maire2016}, 2. \citet{Pecaut2012},
    3. \citet{Macintosh2015}, 4. \citet{Bonnefoy2010},
    5. \citet{Chauvin2005}, 6. \citet{Bonnefoy2014},
    7. \citet{Hinkley2013}, 8. \citet{Jones2016},
    9. \citet{Marois2008}, 10. \citet{Booth2016}, 11. \citet{Wu2017},
    12. \citet{Wolff2017}, 13. \citet{Zhou2014}, 14. \citet{Wu2015a},
    15. \citet{Bowler2015}, 16. \citet{Wu2015b}, 17. \citet{Wu2017b},
    18. \citet{Chauvin2004}, 19. \citet{Chauvin2005b},
    20. \citet{Ricci2017b}, 21. \citet{Song2006},
    22. \citet{Lagrange2009}, 23. \citet{Lagrange2018},
    24. \citet{Bonnefoy2016}, 25. \citet{Su2017},
    26. \citet{Zapata2018}.  \\ $\dagger$ GAIA Data Release
    2. $\dagger\dagger$ Planet temperatures are constrained from
    observations assuming a hot start model.
  \end{flushleft}
\end{table*}

Regular satellites around giant planets are thought to form in
circumplanetary discs (CPD) of gas and dust as by-products of planet
formation \citep{1961SvA.....4..657R, LunineStevenson1982}. In our
Solar system alone, a total of at least 50 regular satellites orbit
the gas giants, including all major moons. These satellite orbits are
close to co-planar with the host planet's equatorial plane, much like
a miniature version of a planetary system. This is true even for
Uranus which has an obliquity of almost 98 degrees. Although the
masses and numbers of regular satellite systems vary from planet to
planet, all regular satellite systems contain a total mass that is
$\sim$10$^{-4}$ times the mass of their central planet. For instance,
there is 2$\times$10$^{-4}$~M$_{\rm Jup}$ worth of regular satellite
mass around Jupiter \citep{CanupWard2006}. The same holds for Saturn
and Uranus\footnote{Neptune's late capture of Triton likely disrupted
  its circumplanetary rings making it an exception}. Is this relation
a common trait for gas giant exoplanetary systems?  It also holds for
close-in rocky planetary systems around late-type stars such as
Trappist~1 \citep[e.g.,][]{Gillon2017}, and in protoplanetary discs
around a variety of stars \citep[e.g.,][]{WilliamsCieza2011}.

In analogy with the minimum mass Solar nebula (MMSN) concept
\citep{Hayashi1981}, a minimum mass subnebula or {\it circumjovian}
disc can be invoked to study the formation of regular satellites, such
as the Galilean moons. In the earlier picture
\citep{LunineStevenson1982, CanupWard2002}, all the ingredients for
the satellites are present from the initial stage. The gases are
eventually dispersed while solids are incorporated into the
satellites.

%% The disc mass surface density, obtained by adding the
%% rock and ice of the Galilean satellites ($\sim$10$^{-4}$~M$_{\rm
%%   Jup}$) with gases to match solar abundances, is about 10$^7~{\rm g\,
%%   cm}^{-2}$ at the planet surface, with a power-law radial falloff.

%% Current exomoon detection efforts rely on indirect methods such as
%% variations in the transit durations \citep{Kipping2009}, with one
%% exomoon candidate recently detected by \citet{Teachey2018}.

Initially, the material within the CPD builds up progressively,
supplied by a near constant inflow of gas and solids from dust traps
in the outer circumstellar disc \citep{CanupWard2002, Cilibrasi2018}.
Three dimensional hydrodynamical simulations agree that CPD radii are
truncated between a third or half of the planet's Hill radius
\citep[e.g.][]{AyliffeBate2009}. \citet{CanupWard2006} found that the
overall properties of satellite systems develop naturally when
satellites grow embedded in these actively supplied CPDs. This model
also reproduces the $\sim$10$^{-4}$ mass fraction of total satellite
mass, as due to a competition between the supply of solids and
satellite loss due to orbital decay in presence of circumplanetary
gas. Upon entering the CPD, the mm-sized dust quickly drifts
inwards. The dust may eventually stall if gas moves radially outwards
in the inner parts of the CPD due to meridional circulation linking
the circumplanetary and protoplanetary discs
\citep{Morbidelli2014}. If so, the mm-sized dust may then form a dense
narrow ring where the streaming instability could set in and form the
km-sized bodies that will later form satellites
\citep{Drazkowska2018}.

The gas in circumstellar discs is thought to dissipate after just a
few million years \citep[e.g.,][]{WilliamsCieza2011}. The details of
what happens next are not well constrained due to lack of
observational evidence. CPDs may remain gravitationally bound to the
planet for a timescale of a few million years to allow enough time to
fully develop Galilean-type satellites
\citep{CanupWard2002}. \citet{Turner2014} reports timescales for full
planet growth of 5-50~Myr, during which the CPD slowly depletes and a
last generation of satellites arises from the {\it protolunar} disc
phase, generalizing the term ``circumjovian'' to a broader
sense. \citet{Crida2012} suggests that moons can still form in massive
primordial rings when they spread beyond the Roche
radius. \citet{KenworthyMamajek2015} reported a system with unusual
transits, which they interpret as circumplanetary rings filling most
of the Hill radius.

In this work, we report on ALMA observations (Sec.~\ref{sec:obs}) of
four directly-imaged exoplanets\footnote{Although some bodies
  considered in this work are in the brown dwarf mass regime, we
  sometimes refer to them as `exoplanets' as this distinction does not
  impact our analysis.} with separations $<$300~au: PZ~Tel~b,
AB~Pic~b, 51~Eri~b, and $\kappa$~And~b. The ages of these systems are
between 21~Myr and 47~Myr, i.e. not much older than the CPD depletion
timescale \citep[$\sim$10~Myr,][]{CanupWard2002} after the dissipation
of the circumstellar gas. These exoplanets are undetected in our
1.3~mm observations but the RMS levels provide upper limits to the
mass reservoir available for satellite formation
(Sec.~\ref{sec:results}). These upper limits are put into context with
previous non detections from the literature
(Sec.~\ref{sec:discussion}).

\section{Data description}
\label{sec:obs}

We used the Atacama Large Millimeter/Submillimeter Array (ALMA) to
observe four directly-imaged exoplanets.  We obtained observations at
1.3 millimeter wavelength using the 12m array with baselines ranging
from 19~metres to up to 3.0~kilometres, resulting in an angular
resolution of $\sim$0\farcs15 (using natural weights), sensitive to
spatial scales of up $\sim$2\farcs0. These observations were performed
during Cycles 3 and 4 (project IDs {\tt 2015.1.01210.S} and {\tt
  2016.1.00358.S}). The cycling time for phase calibration was set to
2~min. Time on science source varies for each target depending on the
exoplanet system properties. On average, each observing block was
integrated for $\sim$38 minutes in total (including calibrations). The
ALMA correlator was configured in Frequency Division Mode (FDM). Three
spectral windows with 2.0~GHz bandwidth were set up for detecting the
dust continuum, centred at 232.469, 217.871, and 215.371~GHz
respectively. A fourth spectral window centred at 230.507 is aimed at
probing for $^{12}$CO gas emission (1.875 GHz in bandwidth with 960
channels). Searching for low-level gas in these systems is beyond the
scope of this paper, and will likely require sophisticated analysis to
produce a detection \citep[e.g., by stacking after correcting by
  Doppler shifts,][]{Marino2016}. Proper motions were accounted for
when more than one observing blocks were used to produce a
concatenated Measurement Set.

All data were calibrated by the ALMA staff using the ALMA Pipeline in
the CASA package \citep{2007ASPC..376..127M}, including offline Water
Vapour Radiometer (WVR) calibration, system temperature correction, as
well as bandpass, phase, and amplitude calibrations.  Image
reconstruction was performed using the CLEAN algorithm (CASA v.5.2,
task {\tt tclean}) with `natural' weights. All spectral windows were
combined in the imaging. No sources were detected in our
observations. The RMS value for each dataset was calculated from the
CLEANed images and are reported in Table~\ref{tab:obs}.

\begin{figure*}
  \centering \includegraphics[width=.84\textwidth]{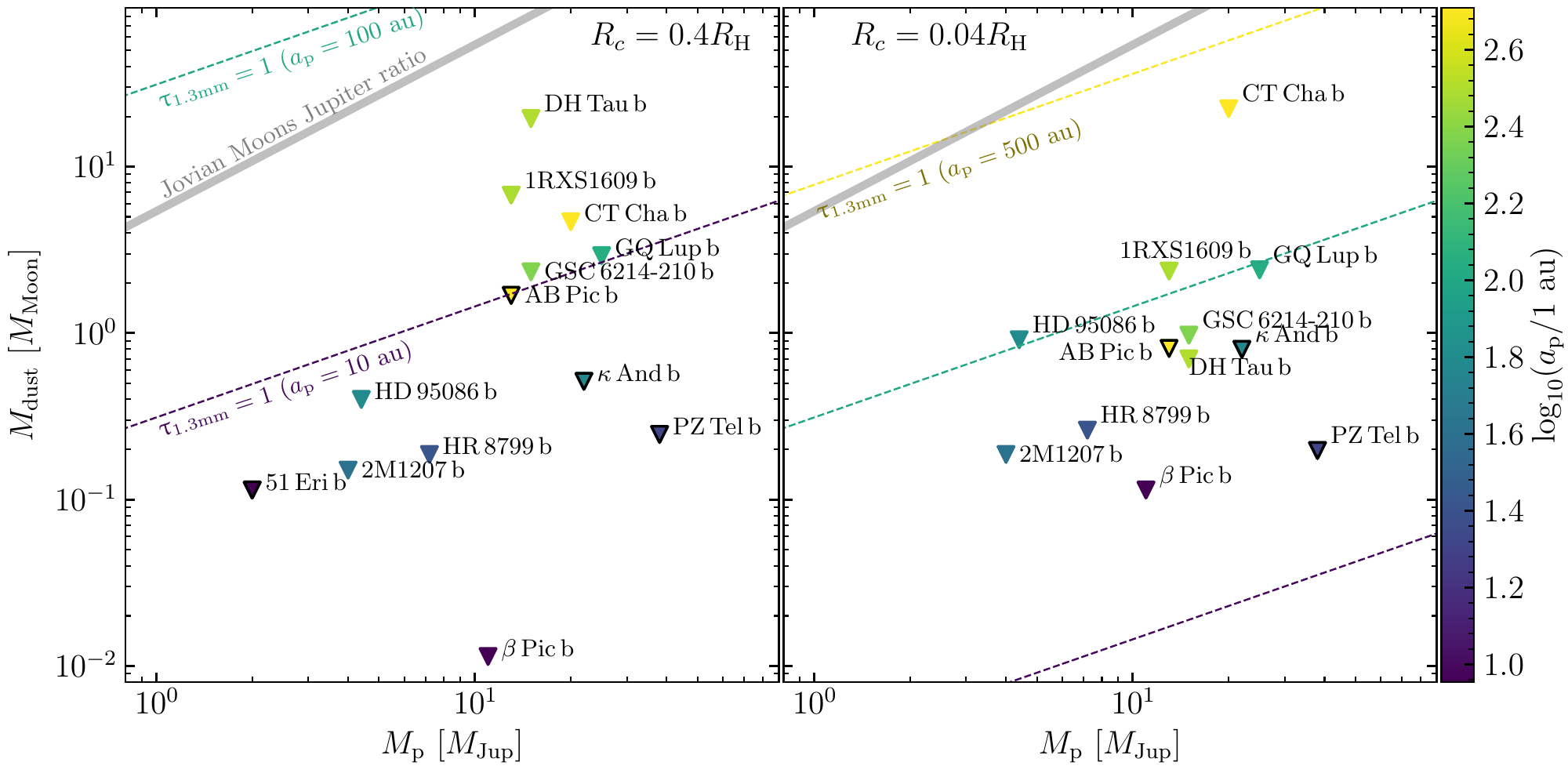}
  \caption{Dust mass upper limits of directly-imaged exoplanets and
    sub-stellar companions observed at 0.89 and 1.3~mm, as a function
    of planet mass. The colour of each marker represents the
    semi-major axis of each planet, which is estimated based on the
    deprojected separation. The four new ALMA observations are
    highlighted with black edges. Calculations are shown for two
    protolunar disc sizes: cutoff radii of 0.4 (left) and 0.04 (right)
    times the planet's Hill radius. The dust masses at which the discs
    become optically thick in the vertical direction at 1.3 mm are
    represented by dashed lines for different planet semi-major
    axes. 51~Eri~b does not feature in the compact disc case as it is
    highly optically thick and unconstraining. }
  \label{fig:MdustvsMp}
\end{figure*}

\section{CPD models and dust mass upper limits}
\label{sec:results}

For all selected candidates, we modelled the protolunar discs as dusty
discs with a surface density profile equivalent to a self-similar
solution for a viscous accretion disc \citep{LyndenBell1974}. We used
a power-law slope of $-1$ and an exponential cutoff radius $r_{\rm c}$
of 0.4 times the planet's Hill radius ($R_{\rm H}$) as our fiducial
model \citep{AyliffeBate2009}. An alternative model with $r_{\rm
  c}=0.04R_{\rm H}$ (about the size of the Galilean moon system) is
also considered to show how sensitive the derived mass is to the disc
radius. The minimum radius of the grid is $r_{\rm c}/10$. If dust
temperatures rise above the sublimation temperature, the disc inner
radius is increased.  A constant disc vertical aspect ratio of 0.1 is
assumed.\footnote{Dust vertical scale-heights are dependent on Stokes
  number and $\alpha$ viscosity. We tested how our approximation
  compares for the case of PZ~Tel and found that a more realistic
  scale-height only affects the emitted fluxes by $\la$10\% when
  compared to a constant aspect ratio for all species.  } A finite
relative inclination of $2\deg$ between the orbit of each planet and
its CPD is used, motivated by regular satellites in the Solar system
which have orbital inclinations ranging from less than a degree up to
$2\deg$.

The dust is assumed to follow a size distribution from 1~$\mu$m to
1~cm (following a power law exponent of $-3.5$) and it is composed of
a mix of astrosilicates, amorphous carbon and water ice. This yields
an absorption opacity of 2.3\,cm$^2$~g$^{-1}$ at 1.3~mm. We input our
model into the {\sc radmc3d} radiative transfer (RT) code to calculate
the total flux expected at 1.3~mm \citep{Dullemond2012}. Both dust
thermal emission and scattering (anisotropic) are considered in the RT
calculation. The star and planet luminosities are included as energy
sources that set the dust temperature. Finally, in order to derive an
upper limit for the total dust mass, we varied the disc mass until the
3$\sigma$ flux upper limit is matched for each disc. When computing
images and observed fluxes we assumed disc inclinations of $30\deg$
from face on for all systems, except for $\beta$~Pic~b, which we
assumed to be inclined by $85\deg$ and co-planar with the
circumstellar debris disc \citep{Lagrange2018,Matra2019}. We note that
this inclination only makes a difference if the discs are massive
enough to become optically thick in the radial direction at 1.3~mm.
The RMS value for $\beta$~Pic~b is calculated as the thermal noise in
a region devoid of circumstellar emission. We interpret this as an
upper limit since upon subtraction of a parametric model for the disc,
there is no point-like emission above 3$\sigma$ \citep{Matra2019}.

%% ------------------------
%% Results
%% ------------------------

In Table~\ref{tab:obs} and Figure~\ref{fig:MdustvsMp} we present the
dust mass upper limits derived from the RT calculations. The limits
are lower for the closest systems as expected, and the ones with
shorter planet semi-major axis since this makes discs warmer and
brighter for the same dust mass. The lowest of these limits is for
$\beta$~Pic~b because it has the deepest ALMA observations, it is the
closest system from our sample, and has the highest temperature field
since it is the closest to its central star. For all systems, the dust
mass upper limits lie below the expected total mass in solids if
exoplanets have the same mass ratio between the total mass in
satellites and planet ($2\times10^{-4}M_{\rm p}$, where $M_{\rm p}$ is
the mass of the exoplanet, see grey band in
Fig.~\ref{fig:MdustvsMp}). This means that there is not enough mass in
dust to form Jovian-like systems, or that satellites must have already
formed or grown to planetesimal sizes which are transparent at
$\sim$1~mm.

In Fig.~\ref{fig:MdustvsMp} we overlay (dashed lines) the dust masses
at which discs become optically thick (assuming constant surface
densities). We find that for cutoff radii of $0.4\,R_{\rm H}$, all
discs are optically thin at 1.3~mm (left panel). However, if discs are
an order of magnitude smaller in radius, the protolunar discs around
51~Eri~b and $\beta$~Pic~b become optically thick at 1.3~mm. In fact,
the upper limit for 51~Eri~b is higher than the maximum possible flux
of a circumplanetary disc in this system (in the compact case the disc
is highly optically thick). Therefore, our flux upper limit for
51~Eri~b does not place a constraint on the disc mass in the small
disc scenario.

The limits derived here can be compared to results by
\citet{Ricci2017b} that assumed optically thin emission and a single
dust temperature scaling with stellar luminosity. We find limits for
DH~Tau~b, GSC~6214-210~B and GQ~Lup~b that are consistent within a
factor of a few when assuming a large cutoff radius, but much lower
for DH~Tau~b if its disc is compact and thus warmer as the heat from
the planet has a larger impact. For 2M1207~b we find a mass that is a
factor of two lower, which can be explained by the higher temperature
in our models compared to the one used by \citet{Ricci2017b}. In our
RT model both the stellar and planet luminosities are included as
heating sources, thus the large dust grains have temperatures of
$\sim$30\,K (between 1 and 10~au), versus the 8\,K assumed in the
previous study.

\section{Discussion}
\label{sec:discussion}

\begin{figure*}
  \centering \includegraphics[width=.84\textwidth]{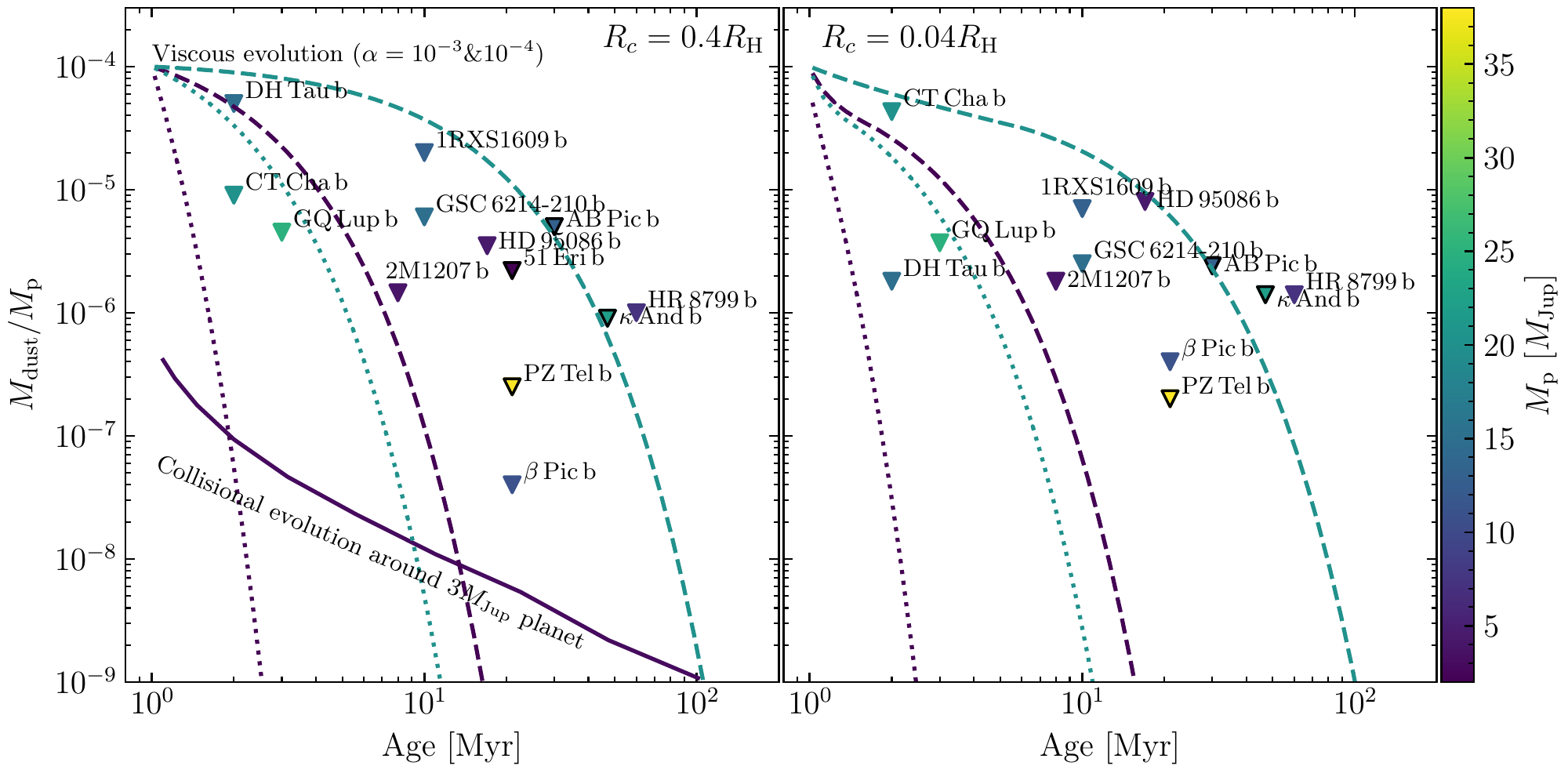}
  \caption{Upper limits on the ratio of dust mass to planet mass for
    directly-imaged exoplanets observed at 0.89 and 1.3 mm as a
    function of stellar age, based on disc models with two cutoff
    radii: 0.4 (left) and 0.04 (right) times the planet Hill
    radius. The dashed and dotted lines represent viscous evolution
    models with $\alpha=10^{-4}$ and $10^{-3}$, respectively, and
    surrounding planets with masses of 2 (purple) and 20~$M_{\rm Jup}$
    (green). The continuous purple line represents the dust levels
    expected if mm-sized dust was created in a collisional cascade in
    a debris disc around a $3 M_{\rm Jup}$ planet.}
  \label{fig:Mdustvsage}
\end{figure*}

The upper limits derived in the previous section are put into context
of two simple evolution models: 1) viscous evolution of a gas disc,
and 2) collisional evolution of a planetesimal disc replenishing a
population of mm-sized dust. These provide order of magnitude
estimates for the expected dust mass in circumplanetary discs as a
function of age, planet and stellar mass, and semi-major axis. 

\subsection{Viscous evolution}
For the viscous evolution scenario we assume an $\alpha$ disc model
where each planet is initially surrounded by an accretion disc rich in
gas and dust (gas-to-dust ratio of 100), with an initial solid mass of
$10^{-4} M_{\rm p}$, where dust particles are smaller than the
molecular mean-free path, i.e. the Epstein regime, where dust and gas
are well coupled (this might not be valid for low gas densities as
large dust grains experience radial drift due to aerodynamics
drag). The dimensionless stopping time or Stokes number (${\rm St}$)
is proportional to the dust size over the gas surface density, for a
given grain internal density. For a vertically isothermal disc, ${\rm
  St}$ can be written as
\begin{equation}
  \rm{St}\!\approx\!0.1\left(\frac{a}{{\rm 1{\scriptscriptstyle mm}}} \right)
  \left(\frac{M_{\rm \scriptscriptstyle dust}/M_{\rm p}}{10^{-4}}
  \right)^{\!-1}\!\!\!\left(\frac{a_{\rm p}}{50\,\rm{au}}
  \right)^{\!2}\! \left(\frac{M_{\rm p}/M_\star}{10^{-2}}
  \right)^{2/3}\!\!\left(\frac{r_{\rm c}}{0.4 R_{\rm {\scriptscriptstyle H}}}
  \right)^{\!2}\!\!,
\label{eq:st}
\end{equation}
\noindent where we assume a grain internal density of 1~g~cm$^{3}$.
Millimetre-sized dust would only be well coupled to the gas (${\rm
  St}<<1$) if disc masses are $\ga$10$^{-2}M_{\rm p}$ or discs have
$r_{\rm c}\,\la\,0.4R_{\rm H}$. Otherwise, large dust grains
containing the bulk of the solid mass will quickly migrate inwards.
Therefore the model presented here provides only an upper limit for
the expected dust masses if the only loss process is the viscous
accretion.

For the viscous evolution we assumed an $\alpha$-disc model
\citep{LyndenBell1974} where viscosity scales linearly with radius,
such that:
\begin{equation}
  \Sigma(r, \widetilde T) = \Sigma_0 \left(\frac{r}{r_0}\right)^{-1} \exp\left[ -\left(\frac{r}{r_0 \widetilde T}\right)\right] \widetilde T^{-3/2},
\end{equation}
where $\widetilde T=(t/t_{\nu,0}+1)$ and $t_{\nu,0}=r_0^2/(3\nu)$ is
the viscous timescale evaluated at $r_0$ (the initial cutoff
radius). The CPD viscous evolution is calculated after the
protoplanetary disc starts dissipating (we assume a dissipation
timescale of 1~Myr). For the disc viscosity we used equation 18 from
\citet{Zhu2018}, assuming a disc temperature of 40~K (this value is
consistent with the temperature at 1~au from the planet in our
models). Then, the disc surface density and mass should decrease with
time as $t^{-3/2}$ and $t^{-1/2}$, with the gas cutoff radius
increasing linearly with time as is the case for a freely expanding
disc. However, this spreading does not happen perpetually since the
CPD must be truncated. Here, we choose a truncation radius $r_{\rm
  t}=0.4R_{\rm H}$, motivated by 3D hydrodynamic simulations of CPDs
\citep[e.g.,][]{AyliffeBate2009}, and assume that at $t=1$~Myr the gas
cutoff radius is already equal to $r_{\rm t}$.  Therefore, for the
evolution after 1~Myr instead we used equation~5 in
\citet{Rosotti2018} that describes the exponential decay of a viscous
disc around a companion (in the context of binary interactions). The
characteristic timescale of the evolution is $t_t=16r_{\rm
  t}^2/(3\upi^2 \nu_c)$, where $\nu_c$ is the viscosity evaluated at
$r_{\rm t}$. In this manner then, the gas and dust mass evolve as
\begin{equation}
    M(t)=M_0 \exp\left( -\frac{t-t_c}{t_t} \right).  \label{eq:M}
\end{equation}

In Fig.~\ref{fig:Mdustvsage}, we compare the dust mass upper limits
with the expected viscous evolution of the dust mass as a function of
system's age. Note that the age of these systems is uncertain and
could vary by $\sim$50\%. The dashed and dotted lines represent models
with $\alpha=10^{-4}$ and $10^{-3}$, respectively. The green and
purple lines correspond to models for planet masses of 20 and
2~$M_{\rm Jup}$, respectively. Note that even if we had considered a
smaller initial cutoff radius, the viscous evolution would be very
similar since an initially small disc will spread quickly until
reaching the truncation radius, after which it will decay
exponentially. We find that, for the high $\alpha$ viscosity case, all
our upper limits are above the model predictions by orders of
magnitude for both disc sizes $0.4R_{\rm H}$ and $0.04R_{\rm H}$ (left
and right panels, respectively), except for CT~Cha when $r_0=0.4R_{\rm
  H}$, DH~Tau when $r_0=0.04R_{\rm H}$ and GQ~Lup in both cases. This
is because these systems are the youngest, thus they are more likely
to have a large dust mass in this scenario. In the low $\alpha$ case,
the disc evolution is slower and our upper limits for planets with
masses above $\sim$10~$M_{\rm Jup}$ lie below the model curves. On the
other hand, the viscous timescale around lower mass planets is shorter
and thus our model predicts masses which are orders of magnitude lower
than the upper limits. Based on our upper limits we conclude that
either CPDs around the more massive of our companions evolve faster
than models with $\alpha\ga10^{-4}$, or mm-size grains are lost
through radial drift or grow into larger objects with very low
opacities at millimeter wavelengths.

\subsection{Collisional evolution}
An alternative scenario for the presence of dust in circumplanetary
discs is to consider secondary origin dust created from collisions of
planetesimals, i.e. like a debris disc \citep[for a detailed
  discussion see][]{Kennedy2011}. Such a disc will collisionally
evolve and its mass will decrease with time as planetesimals are
ground down to the size of small grains blown out by radiation from
the planet or perturbed to $e>1$ by the stellar radiation
\citep{Burns1979}. In order to compute predictions for the evolution
of the disc mass in dust smaller than 1~cm, we used the model
presented in \citet{Marino2017}. This model requires several input
parameters, e.g. the initial mass in solids, planetesimal strength,
level of stirring. We used standard values used to model circumstellar
debris discs for the planetesimal strength and density, and we set the
rest of the parameters to extremes to obtain the most favourable
condition to detect such disc, i.e. slow evolution and high initial
mass. We assumed a planetesimal mean eccentricity and inclination of
0.01 and 0.005; disc radius of 0.4\,$R_{\rm H}$ with a fractional
width of 0.5 to maximise the volume; planet semi-major axis 30~au; and
a planet mass of 3\,$M_{\rm Jup}$ (at the lower end of the
distribution) such that the orbital frequency and collisional rates
are low. Finally we considered a large maximum planetesimal size of
100~km to slow down the collisional evolution, and an extreme initial
solid mass of $10^{-3}M_\mathrm{p}$ (or $\sim10^{-7}M_\mathrm{p}$ in
mm-sized dust assuming a size distribution with an exponent of -3.5)
to estimate the maximum mass of collisionally produced dust that could
be present.

In the left panel of Fig.~\ref{fig:Mdustvsage} the evolution of the
dust mass in a massive planetesimal disc is shown assuming
planetesimals are stirred at 1~Myr (purple line). The dust mass
predicted by this model is orders of magnitude lower than our upper
limits. Higher planet masses, smaller discs in units of $R_{\rm H}$,
higher planetesimal eccentricities or inclinations would make this
evolution even faster. Therefore we do not expect to observe debris
disc like dust around directly-imaged planets, unless the solid mass
is being replenished (e.g. from a circumstellar planetesimal belt) or
large amounts of dust were recently produced is a stochastic collision
between moon embryos.

\begin{figure}
  \centering \includegraphics[width=\columnwidth]{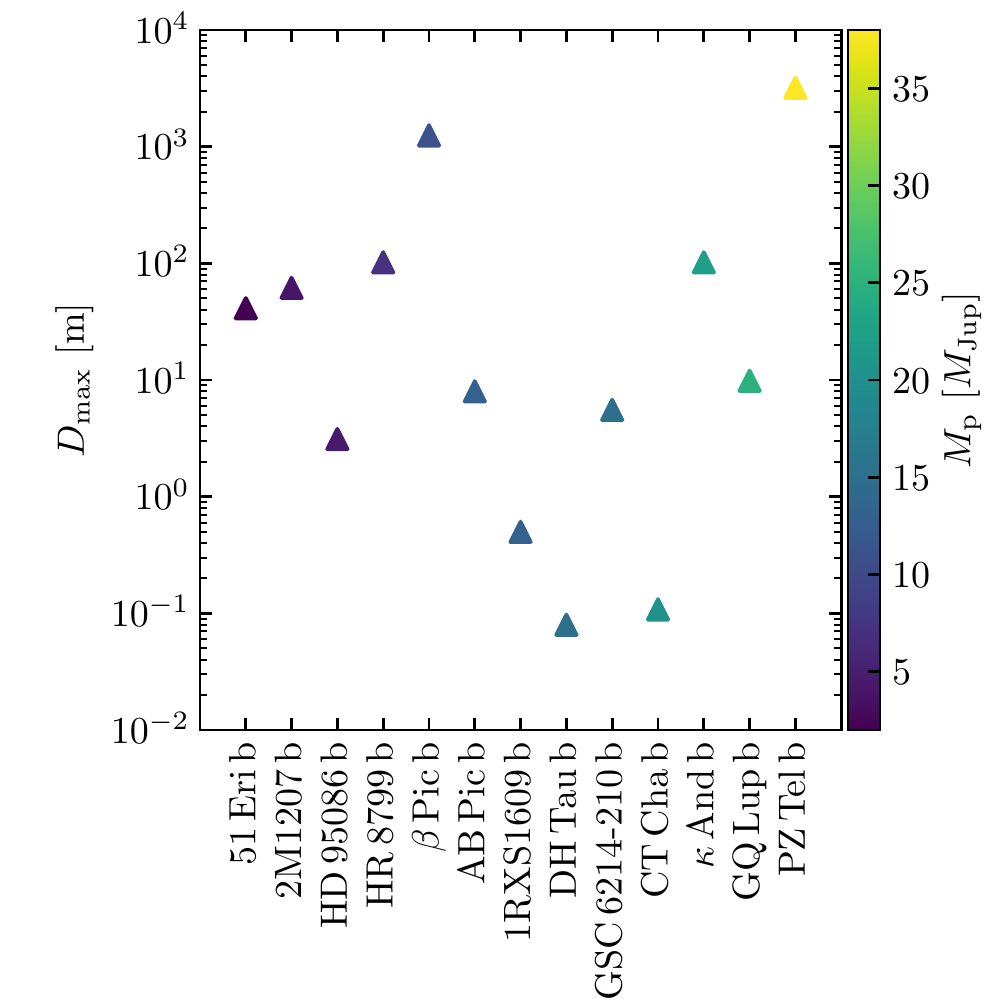}
  \caption{Lower limits on the maximum diameter of solids around each
    directly-imaged companion based on our dust upper limits and
    assuming a solid size distribution with an exponent of -3.5 and a
    total solid mass of $10^{-4}~M_\mathrm{p}$. Sources are sorted by
    planet mass, with the less massive companions on the left, and the
    most massive on the right.}
  \label{fig:amax}
\end{figure}

\subsection{Maximum grain sizes}

Given the derived upper limits on the solid mass in the form of dust
smaller than 1~cm presented in Section~\ref{sec:results}, and by
assuming a size distribution and a total mass of solids, we can derive
a lower limit on the maximum size of particles in these
circumplanetary or protolunar discs. In particular, here we assume a
power law size distribution with an exponent of $-3.5$ and a total
solid mass equivalent to $10^{-4}M_\mathrm{p}$. The solid size lower
limits are then simply
  \begin{equation}
    D_{\max}=2\ \mathrm{cm} \left(\frac{10^{-4} M_\mathrm{p}}{M_\mathrm{dust}}\right)^2,
  \end{equation}
where $D_{\max}$ is the diameter of the largest object in the disc.
In Figure~\ref{fig:amax} we show lower limits on $D_{\max}$ based on
the derived upper limits on $M_\mathrm{dust}$\footnote{We consider the
  largest upper limit on $M_\mathrm{dust}$ between both dust disc
  sizes (0.4~$R_{\rm H}$ and 0.04~$R_{\rm H}$).}. The non detections
translate to a value of $D_{\rm max} >$1~metre for most
sources. Therefore, we conclude that dust growth beyond metre sizes
must happen in timescales $\lesssim10$~Myr. After this timescale, the
solid material is likely in {\em moonetesimal} form.
 
The working hypothesis that motivated the observations and the
analysis presented here assumed that satellite formation is a common
occurrence. Although common in the Solar System, the lack of detection
of circumplanetary material around young exoplanets could also imply
that these planets do not form satellites.

%% ------------------------
%% Conclusions
%% ------------------------
\section{Concluding remarks}
\label{sec:conclusions}

We presented ALMA observations of four young systems host to
directly-imaged exoplanets and sub-stellar companions. Upper limits to
the mass in $\la$mm dust around these planets have been drawn from the
RMS levels using detailed RT modelling.  All upper limits fall below
the expected amount of dust and debris required to explain regular
satellite systems ($\sim$10$^{-4}$\,$M_{\rm p}$).  We interpret this
as likely evidence for rapid growth of dust into $\ga$1~metre
planetesimals (or rather {\em moonetesimals}), which lowers the
protolunar disc opacity at mm wavelengths.

The detection of protolunar discs in dust emission at radio
wavelengths needs deep observations of the youngest exoplanets (e.g.,
DH Tau, CT Cha, GQ Lup) that reach several times higher sensitivities
than current reported observations.

Ongoing observations of dust around accreting planets still embedded
in their protoplanetary discs (e.g., HD\,100546, \citealt{Perez2019},
and PDS\,70, \citealt{Christiaens2019, Isella2019}) could provide
further constraints on the timescales for satellites formation.

%% ------------------------
%% Acknowledgements
%% ------------------------
\section*{Acknowledgements}

We thank the anonymous referee for their constructive comments.  We
acknowledge support from the Millennium Science Initiative (Chile)
through grant RC130007, and CONICYT-FONDECYT grant numbers 1171624,
1171246 and 1191934. S.P acknowledges support from the Joint Committee
of ESO and the Government of Chile.  This work used the Brelka
cluster, financed by Fondequip project EQM140101, hosted at DAS/U. de
Chile. A.Z. acknowledges support from the CONICYT-PAI, Convocatoria
nacional subvenci\'on a la instalaci\'on en la academia, convocatoria
2017, Folio PAI77170087.

\bibliographystyle{mnras}

\begin{thebibliography}{99}

\bibitem[Ayliffe \& Bate(2009)]{AyliffeBate2009} Ayliffe, B.~A., \& Bate, M.~R.\ 2009, \mnras, 397, 657

\bibitem[Bonnefoy et al.(2010)]{Bonnefoy2010} Bonnefoy, M., Chauvin, G., Rojo, P., et al.\ 2010, \aap, 512, A52

\bibitem[Bonnefoy et al.(2014)]{Bonnefoy2014} Bonnefoy, M., Currie, T., Marleau, G.-D., et al.\ 2014, \aap, 562, A111 

\bibitem[Bonnefoy et al.(2016)]{Bonnefoy2016} Bonnefoy, M., Zurlo, A., Baudino, J.~L., et al.\ 2016, \aap, 587, A58 

\bibitem[Booth et al.(2016)]{Booth2016} Booth, M., Jord{\'a}n, A., Casassus, S., et al.\ 2016, \mnras, 460, L10

\bibitem[Bowler et al.(2015)]{Bowler2015} Bowler, B.~P., Andrews, S.~M., Kraus, A.~L., et al.\ 2015, \apjl, 805, L17 

\bibitem[Burns et al.(1979)]{Burns1979} Burns, J.~A., Lamy, P.~L., \& Soter, S.\ 1979, \icarus, 40, 1 

\bibitem[Canup \& Ward(2006)]{CanupWard2006} Canup, R.~M., \& Ward, W.~R.\ 2006, \nat, 441, 834

\bibitem[Canup \& Ward(2002)]{CanupWard2002} Canup, R.~M., \& Ward, W.~R.\ 2002, \aj, 124, 3404

\bibitem[Chauvin et al.(2004)]{Chauvin2004} Chauvin, G., Lagrange, A.-M., Dumas, C., et al.\ 2004, \aap, 425, L29 

\bibitem[Chauvin et al.(2005)]{Chauvin2005} Chauvin, G., Lagrange, A.-M., Zuckerman, B., et al.\ 2005, \aap, 438, L29 

\bibitem[Chauvin et al.(2005)]{Chauvin2005b} Chauvin, G., Lagrange, A.-M., Dumas, C., et al.\ 2005, \aap, 438, L25 

\bibitem[\protect\citeauthoryear{Christiaens, et al.}{2019}]{Christiaens2019} Christiaens V., et al., 2019, ApJ, 877, L33
  
\bibitem[Cilibrasi et al.(2018)]{Cilibrasi2018} Cilibrasi, M., Szul{\'a}gyi, J., Mayer, L., et al.\ 2018, \mnras, 480, 4355
  
\bibitem[Crida \& Charnoz(2012)]{Crida2012} Crida, A., \& Charnoz, S.\ 2012, Science, 338, 1196 

\bibitem[\protect\citeauthoryear{Dr{\k{a}}{\.z}kowska \& Szul{\'a}gyi}{2018}]{Drazkowska2018} Dr{\k{a}}{\.z}kowska J., Szul{\'a}gyi J., 2018, ApJ, 866, 142

\bibitem[Dullemond et al.(2012)]{Dullemond2012} Dullemond, C.~P., et al.\ 2012, Astrophysics Source Code Library, ascl:1202.015 

\bibitem[Gillon et al.(2017)]{Gillon2017} Gillon, M., Triaud, A.~H.~M.~J., et al.\ 2017, \nat, 542, 456 

\bibitem[Hayashi(1981)]{Hayashi1981} Hayashi, C.\ 1981, Progress of Theoretical Physics Supplement, 70, 35

\bibitem[Hinkley et al.(2013)]{Hinkley2013} Hinkley, S., Pueyo, L., Faherty, J.~K., et al.\ 2013, \apj, 779, 153

\bibitem[Isella et al.(2019)]{Isella2019} Isella A., Benisty M., Teague R., Bae J., Keppler M., Facchini S., P{\'e}rez L.~M., 2019, arXiv e-prints, arXiv:1906.06308

\bibitem[Jones et al.(2016)]{Jones2016} Jones, J., White, R.~J., Quinn, S., et al.\ 2016, \apjl, 822, L3

\bibitem[\protect\citeauthoryear{Kennedy, Wyatt, Su \& Stansberry}{2011}]{Kennedy2011} Kennedy G.~M., Wyatt M.~C., Su K.~Y.~L., Stansberry J.~A., 2011, MNRAS, 417, 2281
  
\bibitem[Kenworthy \& Mamajek(2015)]{KenworthyMamajek2015} Kenworthy, M.~A., \& Mamajek, E.~E.\ 2015, \apj, 800, 126

\bibitem[Lagrange et al.(2009)]{Lagrange2009} Lagrange, A.-M., Gratadour, D., Chauvin, G., et al.\ 2009, \aap, 493, L21 

\bibitem[Lagrange et al.(2018)]{Lagrange2018} Lagrange, A.-M., Boccaletti, A., Langlois, M., et al.\ 2018, arXiv:1809.08354 

\bibitem[Lunine \& Stevenson(1982)]{LunineStevenson1982} Lunine, J.~I., \& Stevenson, D.~J.\ 1982, \icarus, 52, 14

\bibitem[Lynden-Bell \& Pringle(1974)]{LyndenBell1974} Lynden-Bell, D., \& Pringle, J.~E.\ 1974, \mnras, 168, 603 

\bibitem[Macintosh et al.(2015)]{Macintosh2015} Macintosh, B., Graham, J.~R., Barman, T., et al.\ 2015, Science, 350, 64 

\bibitem[Maire et al.(2016)]{Maire2016} Maire, A.-L., Bonnefoy, M., Ginski, C., et al.\ 2016, \aap, 587, A56 

\bibitem[\protect\citeauthoryear{Marino, et al.}{2016}]{Marino2016} Marino S., et al., 2016, MNRAS, 460, 2933

\bibitem[Marino et al.(2017)]{Marino2017} Marino, S., Wyatt, M.~C., Kennedy, G.~M., et al.\ 2017, \mnras, 469, 3518 

\bibitem[Marois et al.(2008)]{Marois2008} Marois, C., Macintosh, B., Barman, T., et al.\ 2008, Science, 322, 1348 

\bibitem[Matr{\`a} et al.(2019)]{Matra2019} Matr{\`a}, L., Wyatt, M.~C., et al.\ 2019, arXiv e-prints , arXiv:1902.04081.

\bibitem[McMullin et al.(2007)]{2007ASPC..376..127M} McMullin, J.~P.,  Waters, B., Schiebel, D., Young, W., \& Golap, K.\ 2007, Astronomical Data Analysis Software and Systems XVI, 376, 127

\bibitem[Morbidelli et al.(2014)]{Morbidelli2014} Morbidelli, A., Szul{\'a}gyi, J., Crida, A., et al.\ 2014, \icarus, 232, 266.

\bibitem[Pecaut et al.(2012)]{Pecaut2012} Pecaut, M.~J., Mamajek, E.~E., \& Bubar, E.~J.\ 2012, \apj, 746, 154 

\bibitem[\protect\citeauthoryear{P{\'e}rez, et al.}{2019}]{Perez2019} P{\'e}rez S., et al., 2019, arXiv e-prints, arXiv:1906.06305

\bibitem[\protect\citeauthoryear{Ricci et al.}{2017}]{Ricci2017a} Ricci L., Rome H., Pinilla P., et al.\ 2017, ApJ, 846, 19

\bibitem[\protect\citeauthoryear{Ricci et al.}{2017}]{Ricci2017b} Ricci L., et al., 2017, AJ, 154, 24

\bibitem[Rosotti \& Clarke(2018)]{Rosotti2018} Rosotti, G.~P., \& Clarke, C.~J.\ 2018, \mnras, 473, 5630 

\bibitem[Ruskol(1961)]{1961SvA.....4..657R} Ruskol, E.~L.\ 1961, \sovast, 4, 657 
  
\bibitem[Song et al.(2006)]{Song2006} Song, I., Schneider, G., Zuckerman, B., et al.\ 2006, \apj, 652, 724 

\bibitem[Su et al.(2017)]{Su2017} Su, K.~Y.~L., MacGregor, M.~A., Booth, M., et al.\ 2017, \aj, 154, 225 

\bibitem[Turner et al.(2014)]{Turner2014} Turner, N.~J., Lee, M.~H., \& Sano, T.\ 2014, \apj, 783, 14

\bibitem[Williams \& Cieza(2011)]{WilliamsCieza2011} Williams, J.~P., \& Cieza, L.~A.\ 2011, \araa, 49, 67

\bibitem[Wolff et al.(2017)]{Wolff2017} Wolff, S.~G., M{\'e}nard, F., Caceres, C., et al.\ 2017, \aj, 154, 26 

\bibitem[Wu et al.(2015)]{Wu2015a} Wu, Y.-L., Close, L.~M., Males, J.~R., et al.\ 2015, \apj, 801, 4 

\bibitem[Wu et al.(2015)]{Wu2015b} Wu, Y.-L., Close, L.~M., Males, J.~R., et al.\ 2015, \apjl, 807, L13 

\bibitem[\protect\citeauthoryear{Wu et al.}{2017}]{Wu2017} Wu Y.-L., Close L.~M., Eisner J.~A., Sheehan P.~D., 2017, AJ, 154, 234

\bibitem[Wu et al.(2017)]{Wu2017b} Wu, Y.-L., Sheehan, P.~D., Males, J.~R., et al.\ 2017, \apj, 836, 223 

\bibitem[\protect\citeauthoryear{Wu \& Sheehan}{2017}]{WuSh2017} Wu Y.-L., Sheehan P.~D., 2017, ApJ, 846, L26

\bibitem[Zapata et al.(2018)]{Zapata2018} Zapata, L.~A., Ho, P.~T.~P., \& Rodr{\'{\i}}guez, L.~F.\ 2018, \mnras, 476, 5382 

\bibitem[Zhou et al.(2014)]{Zhou2014} Zhou, Y., Herczeg, G.~J., Kraus, A.~L., et al.\ 2014, \apjl, 783, L17 

\bibitem[Zhu et al.(2018)]{Zhu2018} Zhu, Z., Andrews, S.~M., \& Isella, A.\ 2018, \mnras, 479, 1850 

\end{thebibliography}

%% \bsp
\label{lastpage}

\end{document}